\documentclass[a5paper]{scrartcl}

\usepackage[american]{babel}

\usepackage{graphicx}
\graphicspath{{figures/}}
\usepackage{grffile}

\usepackage{amsmath,amssymb,mathtools}
\usepackage{derivative}
\usepackage{bookmark}
\usepackage[capitalise]{cleveref}
\usepackage{booktabs}
\usepackage{multirow}
\usepackage{siunitx}
\usepackage[frozencache]{minted}
\usepackage{tikz}
\usetikzlibrary{arrows.meta, calc, positioning, shapes.geometric}
\usepackage[labelfont=bf]{caption}
\usepackage{xcolor}

\definecolor{darkblue}{rgb}{0.0, 0.0, 0.75}

\usepackage{adjustbox}

\usepackage{libertine}
\usepackage[libertine]{newtxmath}
\usepackage[scale=0.75]{cascadia-code}
\usepackage[T1]{fontenc}
\usepackage{microtype}

\hypersetup{
  pdftitle = {Improving Interoperability in Scientific Computing
    via MaRDI Open Interfaces},
  pdfauthor = {Dmitry I. Kabanov, Stephan Rave, Mario Ohlberger},
  colorlinks = true,
  linkcolor = darkblue,  
  urlcolor = darkblue,   
  citecolor = darkblue,  
}

\newcommand{\R}{\mathbb{R}}

\renewcommand{\bar}{\overline}

\newcommand{\headerWithBreaks}[2]{{%
\multirow{2}{*}{%
  \begin{minipage}[t]{#1}{#2}\end{minipage}
}
}}

\usepackage[natbib=false, style=numeric, giveninits=true, uniquename=init]{biblatex}
\IfFileExists{./bibliography-external.bib}{%
	\addbibresource{bibliography-external.bib}
}{%
	\addbibresource{bibliography.bib}
}
\newcommand{\citet}[1]{\cite{#1}}
\newcommand{\citep}[1]{\cite{#1}}

\begin{document}
\title{Improving Interoperability in~Scientific Computing via~MaRDI~Open~Interfaces}
\author{%
Dmitry I.\ Kabanov\footnote{Corresponding author, \url{mailto:dmitry.kabanov@uni-muenster.de}},
Stephan Rave,
Mario Ohlberger\\
{\small Mathematics Münster, University of Münster, Germany}}
\date{30 September 2025}
\maketitle

\subsection*{Abstract}

  \emph{MaRDI Open Interfaces} is a~software project aimed at improving
  reuse and interoperability in Scientific Computing by alleviating
  the difficulties of crossing boundaries
  between different programming languages,
  such as C, Julia, or Python,
  in which numerical packages are usually implemented,
  and of switching between multiple implementations
  of the same mathematical problem,
  for example, between different solvers for time integration
  of differential equations.
  The software consists of a set of formal interface specifications
  for common Scientific Computing tasks,
  as well as a~set
  of loosely coupled libraries
  that facilitate implementing these interfaces
  or adapting existing implementations for multiple programming languages
  and handle data marshalling automatically without sacrificing performance,
  enabling users to use different implementations without significant
  coding efforts.
  The software has high reuse potential,
  especially for the tasks such as benchmarking several solvers
  or when low-level access to the interface of a particular solver
  is not required.

\subsection*{Keywords}

unified solver interfaces; scientific computing interoperability; numerical software; scientific software; scientific computing; computational science; cross-language computations; automatic data exchange; data marshalling; programming interface; reusability; foreign function interface

\section{Introduction}%
\label{sec:intro}

The scientific-computing community uses numerical solvers implemented
in different programming languages, such as C, C++, Julia, Python, or R.
In addition to that, each solver has its own programming interface
in terms of function names, order of arguments, and the order of function
invocation.
Due to these two factors, interoperability is inhibited,
as switching from, say, a solver for initial-value problems
for ordinary differential equations from the \emph{SciPy}
package~\citep{Virtanen2020} to another solver implemented in C,
such as the \emph{SUNDIALS} suite~\citep{GardnerEtAl2022, Hindmarsh2005},
requires significant effort from computational scientists.
First, bindings between
two languages must be written for a particular numerical library,
which is a nontrivial task,
even when software packages
such as \emph{Cython}~\citep{Seljebotn2009} or \emph{pybind11}~\citep{Wenzel2017}
can decrease the required workload
by partially automating the procedure.
However, the second factor---different programming interfaces
in different implementations---must
also be addressed, which requires writing additional code and testing.
Moreover, usually such efforts remain coupled to the actual project,
without considering open-sourcing it,
so that multiple computational scientists must redo the same work.

To alleviate these difficulties, we work on software
\emph{Open Interfaces} with two layers.
First, we develop a set of generic interfaces
for typical scientific-computing tasks
(such as integration of ordinary differential equations),
which are used to abstract out discrepancies
between different implementations of these algorithms.
These interfaces are written in a uniform manner
across supported programming languages from a single specification
(currently manually with planned automatic generation in the future).
Second, we develop a mediator library that automates passing data
between different languages
so that writing explicit bindings is not required.
Therefore, computational scientists could switch more easily
from one implementation to another,
while working through the same programming interface,
which leads to faster time-to-solution
and eases the usage of these implementations
for less programming-inclined scientists
as fewer code modifications would be required.

The ideas behind the \emph{Open Interfaces} are
the idea of programming against an interface
and the idea of unifying different numerical solvers under a~single interface
using the \emph{Adapter} pattern
of object-oriented programming~\citep{GammaEtAl1995}
and drive the software design of multiple numerical packages.
Although the ideas themselves are not new---an early and perhaps
the most prominent example in numerics is
\emph{Basic Linear Algebra Subroutines (BLAS)}~\citet{Lawson1979}
published in~1979---
they have gained significant popularity in Scientific Computing
in recent years~\citet{Brown2015}
and drive software design of multiple modern numerical packages.
Some further examples are
the \emph{pyMOR} package by Milk et.~al.~\citet{Milk2016}
for model-order reduction with the Python programming language,
with the algorithms interacting with full-order models
in terms of abstract vector operations
allowing the package users to write new models in other languages
or reuse existing PDE solvers, e.g., \emph{DUNE}~\citep{WellerEtAl1998},
\emph{deal.II}~\citep{ArndtEtAl2021}, \emph{NGSolve}~\citep{GanglEtAl2020}.
Chourdakis et~al.\ \citet{Chourdakis2022} developed library \emph{preCICE}
written in C++ with bindings to other popular languages
that allows users to couple a wide spectrum of finite-elements
and finite-volume solvers, such as \emph{FEniCS}~\citet{BarattaEtAl2023},
and \emph{OpenFOAM}~\citep{BastianEtAl2021}
for multiphysics simulations, providing
facilities for communication, data mapping, and interpolation
between the solvers.
The \emph{Earth System Modeling Framework}~\citep{HillEtAl2004} provides unified
interfaces to different components needed for geophysical
and~weather simulations.
The software package \emph{UM-Bridge}~\citet{
  SeelingerEtAl2023, SeelingerEtAl2025} provides a general way
of decoupling algorithms and models for Uncertainty Quantification problems
via wrapping models in a container and accessing them only via HTTP protocol,
enabling users of languages like C++, Python, R, and Julia to access
models written in different languages.

The main distinction between the aforementioned projects
and~\emph{Open Interfaces} is
that we concentrate on common basic problems in Applied Mathematics
in contrast to, e.\ g., \emph{preCICE} that provides components
in the domain of structural-mechanics and fluid-dynamics simulations.
Another difference is that some of these projects focus on a subset
of languages popular in Scientific Computing:
for example, \emph{pyMOR}'s target language is Python,
while the \emph{Earth System Modeling Framework} provides bindings
only for Fortran and C.
\emph{Open Interfaces}, on the other hand,
strive to treat all programming languages on the user's side as equal,
without forcing the users to use a particular one.

The software package in its current status defines an interface
for solving initial-value problems for ordinary differential equations
(time integration)
and realizes this interface on the user side
for the C and Python programming languages,
with implementations from the \emph{SUNDIALS}, \emph{SciPy},
and \emph{OrdinaryDiffEq.jl} packages.
Already useful in this state,
the package demonstrates what can be achieved in terms of crossing
the language barrier and using different solvers for
the same problem type via the interface for time integration.

For further development,
we plan to define a formal interface specification language
and auto-generate corresponding code
(wrappers for users to invoke implementations, abstract base classes, etc.)
for all supported languages.
We will increase the number of available interfaces and corresponding
supported implementations
with other common numerical tasks, such as optimization.
Also, while we currently concentrate on in-process computations,
where the data are passed from the user to an implementation as pointers,
we also plan to extend the available transport methods
to remote-procedure calls, similar to \emph{UM-Bridge},
for problems, in which the overhead of copying data between processes
is negligible.

\section{Implementation and architecture}\label{sec:desc}

In this section we describe the \emph{Open Interfaces} library,
with the following principles guiding the software architecture.

The main organizational principle is
decoupling of user's code from a numerical implementation
(which can be written in a different language).
The second organizational principle in the architecture is the use
of the C programming language~\citep{KernighanRitchie2002}
for intermediate representation of the data and core libraries
that decouple interfaces from implementations.
Last, the aim to preserve performance as much as possible.
In the following subsections we expand on these principles in details.

\subsection{Architecture, data flow, and realization details}

As was stated before, we decouple users and implementations from each other.
Precisely, we strive to avoid direct pairwise connections between
programming languages on the user side and the implementation side
by inserting a mediator library (which we refer to as \texttt{liboif})
which allows reduce the amount of work required
by computational scientists with different preferred languages, such as C,
Julia, or Python, to use numerical libraries, which can be also written
in different programming languages.

\begin{figure}
  \begin{minipage}{\dimexpr0.50\textwidth-2\tabcolsep}%
    \centering
    \includegraphics[width=0.98\textwidth]{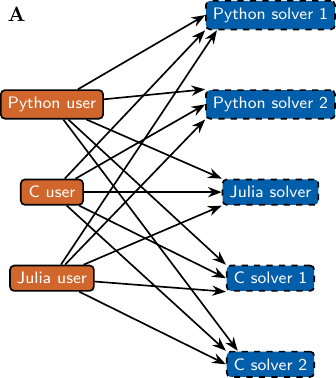}
  \end{minipage}%
  \hfill
  \begin{minipage}{\dimexpr0.5\textwidth-2\tabcolsep}
    \centering
    \includegraphics[width=0.98\textwidth]{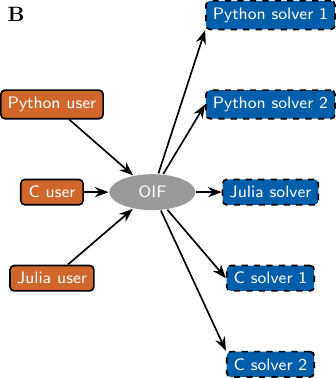}
  \end{minipage}
  \caption{%
    \label{fig:bindings}
    Schematic comparison of two approaches to the problem
    of multiple languages/multiple implementations.
    \textbf{A}~Standard pairwise bindings
    \textbf{B}~Bindings via Open Interfaces (OIF).
  }
\end{figure}

Indeed, as Figure~\ref{fig:bindings} shows schematically,
with $L$ languages and $I$ implementations, one needs
$\mathcal O \left( L \times I \right)$ amount of work
in case of pairwise connections, while only
$\mathcal O \left( L + I\right)$
connections via the mediator library \texttt{liboif}.

\begin{figure}[htbp]
  \centering
  \includegraphics[width=\textwidth]{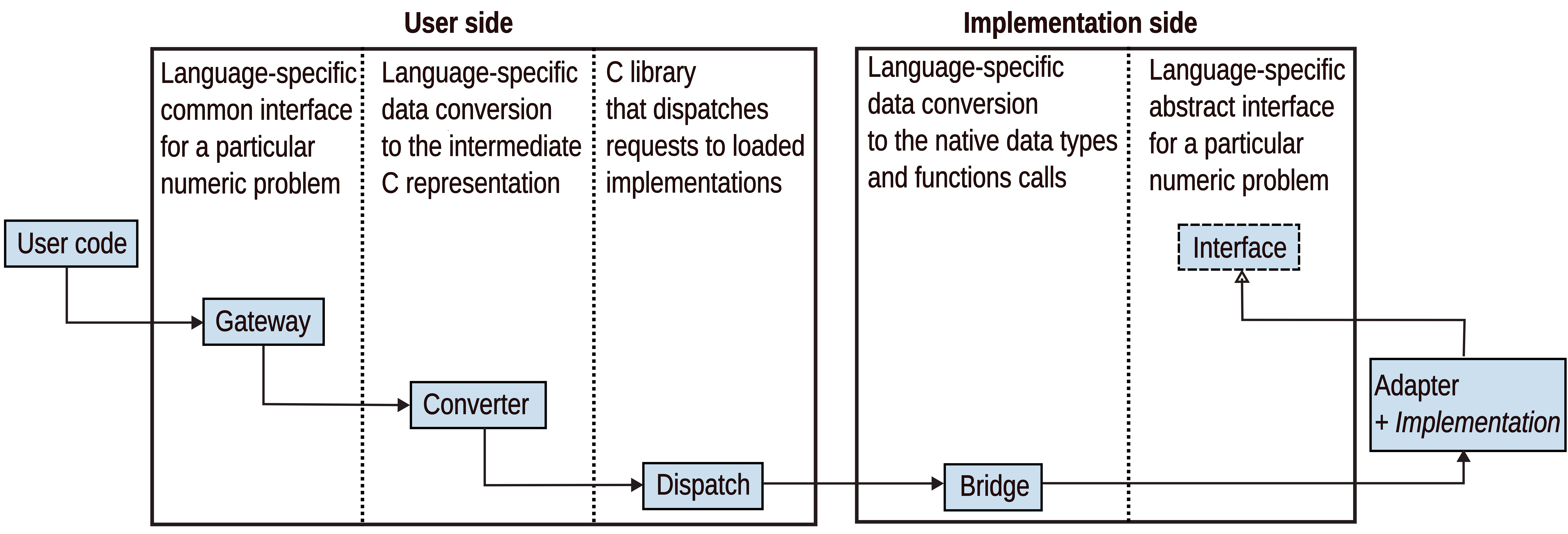}
  \caption{%
    \label{fig:arch}
    Data flow from user to implementation in MaRDI Open Interfaces.
  }
\end{figure}

This library is actually a set of libraries that automate data marshalling
(i.~e., data passing) between different languages and dispatch function calls
from the user to the user-requested implementation.
Figure~\ref{fig:arch} shows the architecture of the library with the data flow
from left to right from the user to a numerical solver.
The vertical lines denote the boundaries between distinct components
of the software organization.

The components on the ``User side'' are the components that users interact
with directly such as an interface for a given numerical problem,
or the components that are stable in a sense that they are belonging
to the library itself and realize operations such as
data conversion and loading implementations.

Components on the ``Implementation side'' are the components that call
implementation methods via provided interfaces.
Particularly, this means that for existing implementations,
an~adapter must be written that translates the calls from \emph{Open Interfaces}
to the calls that the implementations understand.
Formally, such adapters and implementations themselves are not
components of the library.

The responsibilities of each component of the system are the following:

\begin{itemize}
  \item \texttt{Gateway} provides an interface
        for a particular numerical problem,
        through which users can interact with multiple implementations.
        In OOP languages, it also automatically loads/unloads implementations
        in constructor/destructor, while in non-OOP languages it is
        the responsibility of the user.
  \item \texttt{Converter} converts native data types
        to intermediate C representation
        and passes these data further to the \texttt{Dispatch} component.
  \item \texttt{Dispatch} finds the requested implementation on disk,
        reads implementation details, such as the language,
        instantiates the \texttt{Bridge} component for this language
        (if not already instantiated), passes implementation details to it,
        such as module names in Python or shared library name in C,
        and saves the loaded implementation in the table.
  \item \texttt{Bridge} loads an \texttt{Adapter} (along
        with the implementation) from the implementation details,
        calls implementation methods converting data
        from intermediate C representations to the native data types
        for the implementation language, and finally unloads it
        (when requested by the user).
  \item \texttt{Interface} provides an interface that implementation adapters
        must implement.
        This component is an abstract interface
        defining the functions signatures,
        that is, it does not have any functionality itself.
  \item \texttt{Adapter} does not belong to the library but only implements
        the interface described by an \texttt{Interface} component.
        It translates method calls from \emph{Open Interfaces} to the
        implementation.
\end{itemize}

Note that the number of the \texttt{Gateway} and \texttt{Interface} components
is equal to the number of the provided interfaces for numerical problems
multiplied by the number of supported languages:
they must be predefined for each problem type and match each other
(at the present time
we write them by hand with a plan to use automatic generation in the future);
the number of the \texttt{Converter} and \texttt{Bridge} components is equal to
the number of supported programming languages on user and implementation sides,
respectively; the \texttt{Dispatch} component is a singleton that is oblivious
of the languages it connects on both sides
as well as the supported problem types.

To understand the workflow better, let's consider
three main phases of interaction between the user
and \emph{Open Interfaces}.
In \emph{initialization} phase, the user specifies
what interface they want to use along with the string identifier for the
needed implementation.
In the \emph{calling} phase, the user interacts with the
implementation through the \texttt{Gateway} instance, which is a class
in object-oriented languages like Python, or a set of functions
in languages like C and Julia.
When the user is finished using a particular implementation,
this implementation is unloaded in the \emph{unloading} phase.

\begin{figure}[htbp]
  \centering
  \includegraphics[width=\textwidth]{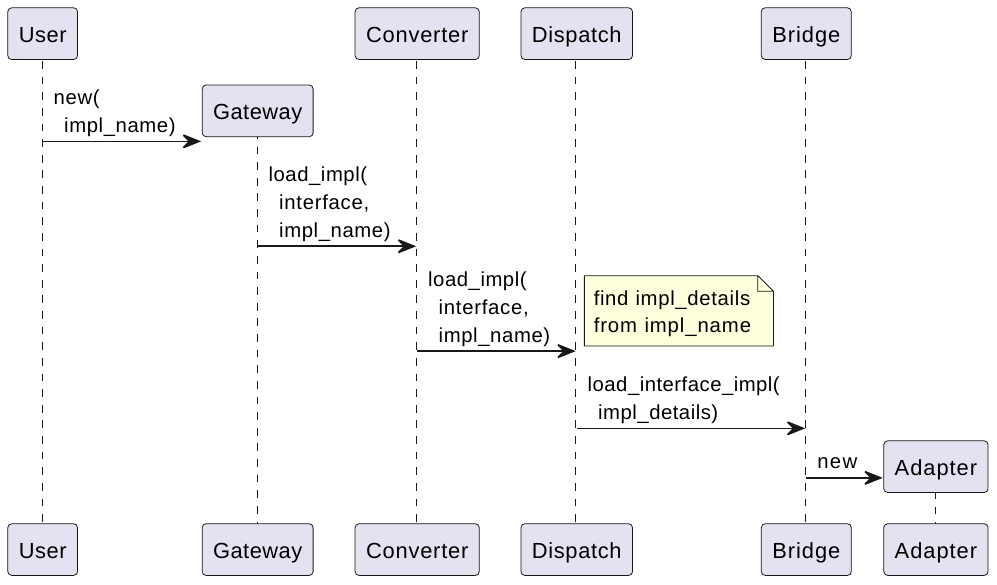}
  \caption{%
    \label{fig:desc:init}
    UML sequence diagram for the initialization phase, in which user
    requests an implementation for an interface of interest.
    The abbreviation ``impl'' stands for ``implementation''.
  }
\end{figure}

\textbf{Initialization phase}.
Figure~\ref{fig:desc:init} shows the UML sequence diagram of function calls
that occur in the initialization phase.
The user creates a \texttt{Gateway} component for a particular interface,
passing a string identifier for the needed implementation of that interface.
\texttt{Gateway} passes information about the interface and the implementation
to the \texttt{Converter}, which converts this information to two C strings.
These strings are passed to the \texttt{Dispatch} component,
which first finds on disk the implementation details from the data pair
``interface-implementation''.
The details are language-specific, so that \texttt{Dispatch} itself
only processes the first chunk of the details---the language---and loads
a~corresponding \texttt{Bridge} component
(if it is not loaded before).
The \texttt{Bridge} component processes the rest of implementation details
(which are names of shared libraries in C, or Python modules, etc.)
and loads the \texttt{Adapter} itself, which
in turn loads the implementation.

At the end of the initialization phase, the user holds an \emph{implementation
  handle} \texttt{implh} (implicitly in object-oriented languages such as Python
or explicitly in languages such as C), which further acts as the identifier
connecting all components together. Additionally, the \texttt{Dispatch}
component saves the initialized implementation in the table
of loaded implementations: the instantiated
\texttt{Bridge} and the loaded implementation,
which can contain multiple pieces of information,
depending on the language, in which it is written.
For example, it can be a loaded shared library in case of C
or instantiated object in Python.
Besides that, for non-OO languages,
abstract state information can be also preserved in this table, which
is determined by a particular implementation's needs.

\begin{figure}[htbp]
  \centering
  \includegraphics[width=\textwidth]{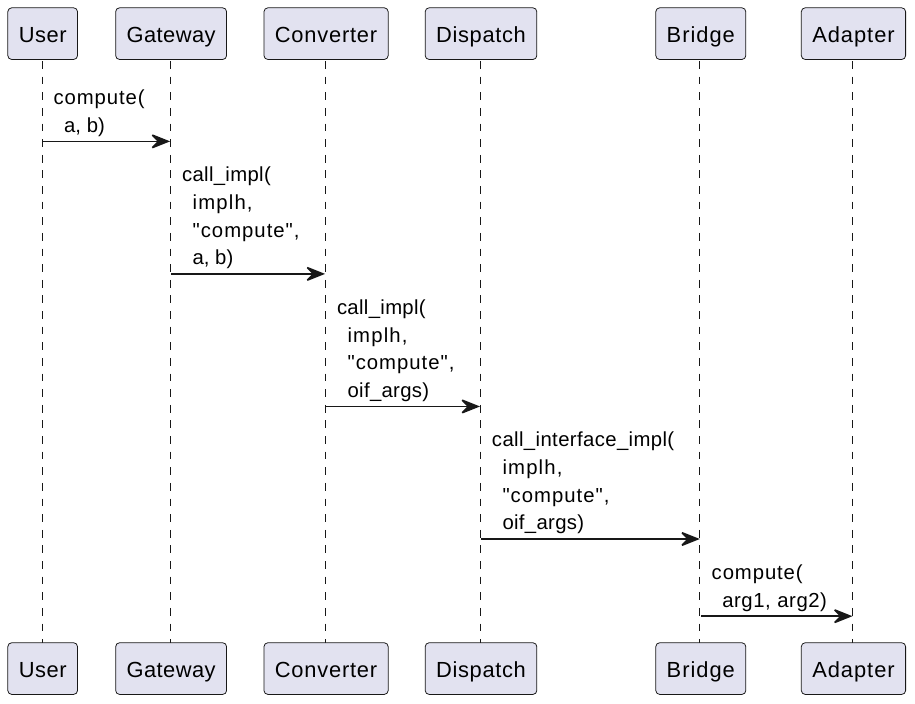}
  \caption{%
    \label{fig:desc:call}
    UML sequence diagram showing the function invocations when the user
    does the actual computations. The diagram shows invocation of a hypothetical
    method \texttt{compute} with two arguments that are converted to a list
    \texttt{oif\_args} and then unpacked by the \texttt{Bridge} component
    to native data types of the implementation.
  }
\end{figure}

\textbf{Calling phase}.
Figure~\ref{fig:desc:call} shows the generic sequence of function
calls through the system when the user does the actual numerical computations.
Each function call from the user to the \texttt{Gateway}
is passed further to the \texttt{Converter}
that packs different arguments into a list with each argument converted
to its C representation and passes the implementation handle,
the interface method name and the packed arguments
to the \texttt{Dispatch} component.
The \texttt{Dispatch} component does here only minimal work: it finds the record
in the implementation table and passes the information further
to the corresponding \texttt{Bridge} component.
The \texttt{Bridge} component unpacks the arguments
from the list and transforms them to the native data types of the language,
and then invokes the requested method on the \texttt{Adapter} component.
The \texttt{Adapter} component does the computations
by using the implementation.

To simplify memory management, we strive to have all necessary memory
allocations done on the user's side,
explicitly by the user or inside the \texttt{Gateway} component.
However, in general this is not always possible, and in such cases
the memory ownership must be explicitly pronounced in interface definitions,
to diminish the possibility of memory leaks.

\begin{figure}[htbp]
  \centering
  \includegraphics[width=\textwidth]{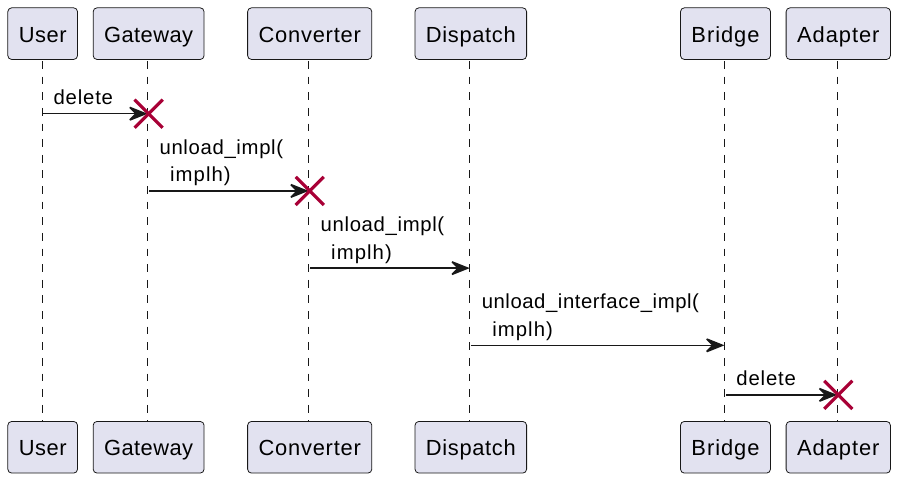}
  \caption{%
    \label{fig:desc:unload}
    UML sequence diagram showing the function invocations
    for the unloading phase when the user has finished
    using the implementation.
  }
\end{figure}

\textbf{Unloading phase}.
When the user has finished using the implementation,
it is unloaded from the memory as shown in the Figure~\ref{fig:desc:unload}.
Precisely, when the user deletes the \texttt{Gateway} component
(in an OO language like Python), the request to unload the implementation
is executed automatically, while it must be executed explicitly by the user
in languages like C.
When the \texttt{Dispatch} component processes this request,
it removes the implementation from its table and releases
all memory related to the implementation.

Note that, as shown in the Figure~\ref{fig:desc:unload},
the \texttt{Dispatch} component is not deleted, as it still continues
to keep the implementation table.
The \texttt{Bridge} is not deleted either as, for example, the embedded
Julia interpreter cannot be started again after finalization.

\subsection{Data types}
As stated before, we use C data types for intermediate representation,
as popular languages used in Scientific Computing,
such as Python and Julia,
have facilities to communicate with C
and making function calls to C,
providing means of conversion of the data from the intermediate
representation to the native data types of these languages.

Particularly, it is easy to convert Python and Julia integer data types to
C~\texttt{int} data type, provided that the integers are representable
in~32~bits.
Also, conversion of binary double-precision floating-point numbers
is straightforward between C and other languages due to the widespread use
of the IEEE 754 standard for floating-point arithmetic~\citep{IEEE754}.

Data marshalling of arrays of double-precision floating-point numbers
is made possible by using an auxiliary data structure \texttt{OIFArrayF64} that,
similarly to NumPy arrays~\citep{HarrisEtAl2020} or Julia arrays,
represents $n$-dimensional array for given
$n \in \mathbb N$ and packs data together with the number of dimensions $n$
and the array shape, that is, the size of the array along each dimension.
This data structure enables a~uniform function signature
among supported languages (currently C, Python, and Julia)
as then the arrays are given as a single
function argument in all these languages (in contrast with traditional use
of C arrays, where data and dimensions are provided as separate arguments).
Correspondingly, we use NumPy C API and Julia C API to convert to
\texttt{OIFArrayF64} and back when needed.

We also support read-only strings that can be used to pass information such
as, e.\ g., a name of an integrator.

As it is common in scientific computing to pass callback functions to numerical
solvers, Open Interfaces support passing functions between different languages.
This is achieved in the following manner.
Additional data structure \texttt{OIFCallback} is used that encodes information
about the original language of the callback function, the function itself
in this language, and the C-compatible version of this function.
Consequently, on the language-specific dispatch level, if the user-facing
and implementation languages are the same, the original callback function
is used to avoid performance penalties, while the C-compatible callback
is wrapped in the programming language of the implementation.

Additionally to the callback, passing a generic memory pointer is supported
which is required, for example, to pass context to the callback functions.
Although in languages like Python one can simply use closures
to pass the context, in languages like C it is the only way to achieve this.

Finally, simple dictionaries of key-value pairs, where keys are strings,
and values are either integer or floats, are supported to pass generic
options that are implementation-specific.

For each supported data type, the data are passed between software components
along with integer identifiers allowing to restore the type on the receiver
end. We use the following symbolic constants
further in the text to refer to the actual data types:
\begin{itemize}
  \item \texttt{OIF\_INT}: 32-bit integers,
  \item \texttt{OIF\_FLOAT64}: 64-bit binary floating-point numbers,
  \item \texttt{OIF\_ARRAY\_F64}: arrays of 64-bit binary floating-point numbers,
  \item \texttt{OIF\_STR}: strings with one-byte characters
  \item \texttt{OIF\_CALLBACK}: callback functions,
  \item \texttt{OIF\_USER\_DATA}: user-data objects of volatile type,
  \item \texttt{OIF\_CONFIG\_DICT}: dictionary of key-value options pairs.
\end{itemize}

It is assumed that each symbolic constant is replaced with the actual data type
when used in a particular language: for example, \texttt{OIF\_ARRAY\_F64}
resolves to the provided data structure \texttt{OIFArrayF64} in C
and to NumPy arrays with \texttt{dtype=numpy.float64} in Python.

\subsection{Data passing and function calls}

Copying data, especially, large arrays, impedes performance,
as modern computer architectures are bounded by memory operations.
Hence, in development of \emph{Open Interfaces}, we avoid copying data and pass
all data as pointers, which makes all conversion operations fast and cheap.
Conversion of integer and floating-point numbers is cheap by itself,
but even for arrays, it is a matter of creating a thin wrapper
around an~actual data pointer.

To invoke functions between different languages,
the \texttt{libffi}\footnote{\url{https://sourceware.org/libffi/}}
library is used, either indirectly, e.\,g.,
using \texttt{ctypes} in Python,
or explicitly for calling C functions dynamically.

Note that we use C convention of functions returning an integer to indicate
an error. When the resultant integer is zero, the function invocation is
successful, and not otherwise. For languages that support exceptions,
an exception is raised on the user side,
so that the user does not have to check every function call for errors.

\subsection{Interface for initial-value problems for ODEs}%
\label{sub:ivp}

One of the use cases already developed in this project is the open interface
for solving initial-value problems (IVP)
for ordinary differential equations (ODEs),
namely, for problems of the form:
\[
  y'(t) = f(t, y), \quad y(t_0) = y_0,
\]
where $y(t) \colon \R \to \R^{n}$, $f(t, y) \colon \R \times \R^{n} \to \R^{n}$,
$y_{0}$ is the initial system state at time $t_{0}$.
We refer to this interface as \texttt{IVP} in the following text.

This interface, in a C-like pseudolanguage, consists of the following
function invocations:
\begin{minted}{C}
// Set an initial condition
OIF_INT set_initial_value(OIF_ARRAY_F64 y0, OIF_FLOAT64 t0);
// Set the right-hand side (RHS) callback function
OIF_INT set_rhs_fn(OIF_CALLBACK rhs_fn);
// (Optional) Set relative and absolute tolerances
OIF_INT set_tolerances(OIF_FLOAT64 reltol, OIF_FLOAT64 abstol);
// (Optional) Set user-defined data that are passed
// to the right-hand side function
OIF_INT set_user_data(OIF_USER_DATA user_data);
// (Optional) Set integrator and its parameters
OIF_INT set_integrator(OIF_STR integrator, OIF_CONFIG_DICT params);
// Integrate to new time and write the solution to vector y
OIF_INT integrate(OIF_FLOAT64 t, OIF_ARRAY_F64 y);
\end{minted}

This interface assumes that the right-hand-side (RHS) function has the following
signature:
\begin{minted}{C}
OIF_INT rhs_fn(
  OIF_FLOAT64 t,
  OIF_ARRAY_F64 y,
  OIF_ARRAY_F64 ydot,
  OIF_USER_DATA user_data
)
\end{minted}
where \texttt{ydot} is the output variable
to which the value of \(f(t, y)\) is written.

\section{Quality control}

\emph{MaRDI Open Interfaces} uses testing and continuous integration
for quality control.
The test suite includes tests for all supported languages on the user's side
and assesses the behaviour of all included implementations
against problems with known analytical solutions,
with additional tests against edge cases such as wrong input.
Further, unit tests for checking internal data structures and functions
are also included.
The test suite is based on standard tools for corresponding languages,
such as \emph{Google Test}
\footnote{\url{https://github.com/google/googletest}} for C
and \emph{pytest}~\cite{KrekelEtAl2004} for Python.

Additionally, we use continuous integration service
provided by \emph{GitHub}
that regularly runs the tests as well as static analyzers and style checkers,
ensuring that the software's results are reproducible and correct
and that the code style remains consistent.

To help users learn the package,
we provide
documentation\footnote{https://mardi4nfdi.github.io/open-interfaces/}
for the most recent release, with instructions on installation,
how to check that the compiled package works as expected by running tests,
and how to get started using the package with several examples.
The full source of these examples is provided in subdirectory \texttt{examples}
of the source code.

\section{Examples}\label{sec:results}
We explain shortly in this section
a couple of examples of using the software
along with the outputs.

\subsection{Solve Burgers' equation from C}

The first example is solving in C inviscid Burgers' equation:
\begin{equation}
  \label{eq:results:burgers}
  \begin{aligned}
     & \pdv{u}{t} + \pdv{\left( u^{2} / 2 \right)}{x} = 0,
    \quad t \in [0, 2], \enspace x \in [0, 2]              \\
     & u(0, x) = 0.5 - 0.25 \sin \left( \pi x \right)      \\
     & u(t, 0) = u(t, 2)
  \end{aligned}
\end{equation}
after converting it to a system of ordinary differential equations:
\begin{equation}
  \label{eq:results:burgers:mol}
  \begin{aligned}
     \odv{U_{i}}{t} & = -\frac12 \frac{U_{i+1/2}^2 - U_{i-1/2}^2}{\Delta\,x}, \\
     U_{i}(0)       & = 0.5 - 0.25 \sin \left( \pi x_{i} \right)              \\
     U_{-1/2}       & = U_{N+1/2},
  \end{aligned}
\end{equation}
where $U(t)$ is a grid function discretized on a finite-volume grid
${x_0,\dots,x_N}$ of resolution $N$.
To solve this system, we approximate fluxes at the finite-volume interfaces
$x_{i-1/2}$, $i=0,\dots,N$, using the global Lax--Friedrichs flux~\citep{LeVeque2007}.
We integrate the system~\eqref{eq:results:burgers:mol} to final time 10.

We use the C version of the \texttt{IVP} open interface for integrating in time,
but for brevity omit checking error status codes here.
First, we initialize a desired implementation (which is in the variable
\texttt{impl}) of the \texttt{IVP} interface:
\begin{minted}{C}
  ImplHandle implh = oif_init_impl("ivp", impl, 1, 0);
\end{minted}
and after the initialization, the user obtains an implementation handle
\texttt{implh} that is used in all subsequent function calls.
Note, that the third and the fourth arguments of the function
\texttt{oif\_init\_impl} are for specifying the version of an implementation,
however, right now this numbers are not used.

Now, we allocate a vector \texttt{y} for the solution of size \texttt{N},
set initial conditions (\texttt{y0} at time \texttt{t0}),
user-provided right-hand side callback \texttt{rhs},
and user data (context) that must be passed to the right-hand side callback
(in this case, spatial step \texttt{dx})---all this information
is derived from Eq.~\eqref{eq:results:burgers:mol}:
\begin{minted}{C}
  OIFArrayF64 *y = oif_create_array_f64(1, (intptr_t[1]){N});

  status = oif_ivp_set_initial_value(implh, y0, t0);
  status = oif_ivp_set_rhs_fn(implh, rhs);
  status = oif_ivp_set_user_data(implh, &dx);
\end{minted}

Finally,
we integrate the system for desired number of steps \texttt{n\_time\_steps}
using uniform time step \texttt{dt}:
\begin{minted}{C}
  for (int i = 0; i < n_time_steps; ++i) {
      t = t0 + (i + 1) * dt;
      status = oif_ivp_integrate(implh, t, y);
  }
\end{minted}

Note that due to the lack of object-orientation in C,
the user must keep the handle to implementation \texttt{implh}
and pass it to every function of the \texttt{IVP} interface.

Figure~\ref{fig:results:solns}A shows the solution
of the problem~\eqref{eq:results:burgers:mol}
obtained using the \texttt{dopri5} integrator (Dormand-Prince 5(4) method)
from the \texttt{scipy\_ode} implementation,
which is an adapter to the \emph{SciPy} package.

The full code of this example can be found in the repository
at the path \texttt{examples/call\_ivp\_from\_c\_burgers\_eq.c}.

\subsection{Solving Van der Pol's equation from Python}
In the second example, we use Python and solve the Van der Pol's oscillator equation:
\begin{equation}
  \label{eq:vdp}
  \odv[order=2]{x}{t} - \mu
  \left(
  1 - x^{2}
  \right) \odv{x}{t} + x = 0, \quad
  x(0) = 2.
\end{equation}
with \(\mu = 1000\).
The problem~\eqref{eq:vdp} can be converted to a system
of first-order ODEs and solved again using the \texttt{IVP} open interface.
However, due to the value of the parameter $\mu$, the system is stiff
and the system must be solved using implicit ODE
solvers~\citet{LeVeque2007}.

\begin{figure}
  \centering
  \includegraphics[scale=0.8]{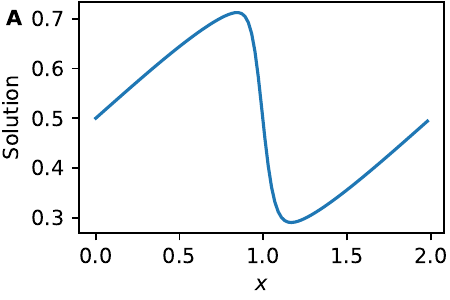}
  \hfill
  \centering
  \includegraphics[scale=0.8]{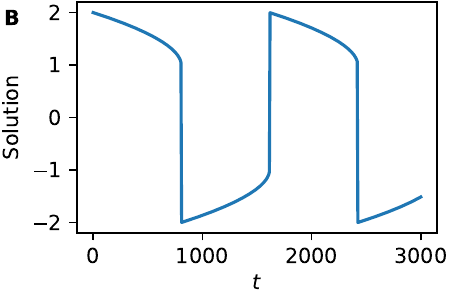}
  \caption{%
    Example solutions obtained using \emph{MaRDI Open Interfaces}:
    \textbf{A}
    Solution of the problem~\eqref{eq:results:burgers:mol}
    using \texttt{scipy\_ode} implementation with integrator \texttt{dopri5}
    (Dormand--Prince 5(4) method).
    \textbf{B} Solution of the problem~\eqref{eq:vdp}
    using \texttt{jl\_diffeq} implementation with integrator \texttt{Rosenbrock23}.
    \label{fig:results:solns}
  }
\end{figure}

Indeed, if we try to solve the system using, for example, \texttt{scipy\_ode}
IVP implementation, which by default uses the Dormand--Prince algorithm~\citet{DormandPrince1980}:
\begin{minted}{Python}
  p = VdPEquationProblem(mu=1000, tfinal=3000)  # User class
  s = IVP("scipy_ode")
  s.set_initial_value(p.y0, p.t0)
  s.set_rhs_fn(p.compute_rhs)
\end{minted}
then the computations fail due to stiffness:
\begin{minted}{Shell}
UserWarning: dopri5: problem is probably stiff
\end{minted}

Switching to the \texttt{jl\_diffeq} implementation that adapts Julia's
\texttt{OrdinaryDiffEq.jl} package to the \texttt{IVP} interface
of Open Interfaces, one can choose
one of the implicit solvers that solve the problem~\eqref{eq:vdp} efficiently:
\begin{minted}{Python}
  p = VdPEquationProblem(mu=1000, tfinal=3000)  # User class
  s = IVP("jl_diffeq")
  s.set_initial_value(p.y0, p.t0)
  s.set_rhs_fn(p.compute_rhs)
  s.set_integrator("Rosenbrock23")
\end{minted}
where \texttt{Rosenbrock23} is a Julia's implementation
of the \textsc{MATLAB}'s \texttt{ode23s} integrator~\citep{ShampineReichelt1997}.
The integrator succeeds,
with the solution shown in Figure~\ref{fig:results:solns}B.

The full code of this example is available in the source code
repository at the path \texttt{examples/call\_ivp\_from\_python\_vdp.py}.

\section{Performance analysis}
To assess the performance loss that might be introduced
by the core components of the \emph{Open Interfaces},
namely, components that convert data and dispatch function calls,
we conduct performance comparison using different combinations
of user languages and implementations.

All run time results in this section are reported as 95~\% confidence intervals
\(\bar{r} \pm 1.96 \mathrm{se}\)
based on a sample of runtimes $r_{1},\dots,r_{n}$, $n=30$,
with sample mean
\[
  \bar{r} = \frac{\sum_{i=1}^{n}r_{i}}{n},
\]
and the standard error of the mean
\[
  \mathrm {se} = \sqrt{\frac{1}{n(n-1)}\sum_{i=1}^{n}\left( r_{i} - \bar{r}\right)^{2}}.
\]

The performance study is based
on time integration of Problem~\eqref{eq:results:burgers:mol}.

The runtime can be heavily affected by the performance
of the used time integrator and the performance
of the RHS evaluation,
and we try to make sure that these two factors are of the same magnitude
between different implementations.

For all implementations we use the Runge–-Kutta 5(4)
method of~\citet{DormandPrince1980} with relative and absolute tolerances
set to \num{e-6} and \num{e-12}, respectively.
The first implementation is in C and is translated from the original code
from~\citet{HairerEtAl1993}, and we refer to it in the further text
as \texttt{DOPRI5-C}.
The second implementation is in Julia, from the \emph{OrdinaryDiffEq.jl}
package, from which we use the integrator \texttt{DP5}.
The third implementation is from Python's \emph{SciPy} package,
from which we use the integrator \texttt{DOPRI5}
(which is actually written in Fortran~\citep{HairerEtAl1993}).

We also make sure that the RHS implementations
have similar performance across all languages.
To do that, we optimize by hand.
Additionally, we compile the C RHS implementation as a shared library using
the Clang compiler~\citep{clang}
and optimize the Python RHS version using Numba~\citep{LamEtAl2015}.
Thus, all versions of the RHS function are compiled to machine code
via LLVM compiler infrastructure\footnote{\url{https://llvm.org/}}.
Table~\ref{tab:results:burgers:rhs} shows the run times, in seconds,
of evaluating RHS implementations \num{10000} times
for $N=\num{6400}$.
As the data show, all RHS implementations have similar performance.
These performance experiments were conducted using a workstation
with an Intel Xeon E5-1630 v3 processor
and 32 gigabytes of memory.

\begin{table}[hbtp]
  \caption{%
    Runtimes, in seconds, of evaluating RHS implementations
    for system~\eqref{eq:results:burgers:mol} \num{10000} times
    at resolution $N=\num{6400}$.%
    \label{tab:results:burgers:rhs}
  }
  \centering
  \begin{tabular}{l c}
    \toprule
    Implementation language & Runtime, seconds \\
    \midrule
    C                       & 0.115 ± 0.008    \\
    Julia                   & 0.122 ± 0.016    \\
    Python (Numba)          & 0.116 ± 0.001    \\
    \bottomrule
  \end{tabular}
\end{table}

Table~\ref{tab:results:burgers:runtimes} shows the results of the performance
study.
We run simulations for three different resolutions
$N\in \{1600, 6400, \num{25600}\}$ and compare different combinations
of user languages and implementations.
The table consists of three blocks corresponding to three comparisons.

\textbf{Comparison 1}.
First we compare performance of time integration using
the implementation \texttt{DOPRI5-C}
invoking it via \emph{Open Interfaces} and directly.
As can be seen,
performance penalty is small as for all used resolutions
the run times do not differ for more than six percents.

\textbf{Comparison 2}.
Here we compare run times of using the \texttt{DP5} implementation
from C via \emph{Open Interfaces} against
using the same \texttt{DP5} implementation from Julia directly.
In this comparison, the first two cases use the C RHS implementation,
while in the third case the Julia RHS implementation.
We can see that for $N=1600$ there is a performance
penalty of 20--50~\% because of invoking \texttt{DP5} from C,
that is, for this resolution,
the overhead of \emph{Open Interfaces} is non-negligible.
However, already for moderate resolution $N=6400$
the difference in runtime is not more than 5~\%,
and for $N=\num{25600}$ it is less than 2~\%.
Between the first and the third cases in this comparison,
the first case, with the C RHS via \emph{Open Interfaces}, is slightly faster
than the third case, with the Julia RHS and direct invocation of the \texttt{DP5}
solver, which is expected, as the C RHS is slightly faster itself
than the Julia RHS (see Table~\ref{tab:results:burgers:rhs}).

\textbf{Comparison 3}.
In this comparison we consider user code implemented in Python,
with the RHS functions optimized via Numba, and run
two implementations---\emph{DOPRI5} from \emph{SciPy} and \texttt{DP5}
from \emph{OrdinaryDiffEq.jl}---via Open Interfaces and directly.
We can see that there is practically no runtime difference
between using the \texttt{DOPRI5} implementation from Python codes
directly or via \emph{Open Interfaces}.
If the \texttt{DOPRI5} implementation from SciPy is replaced
with the integrator \texttt{DP5} from the \emph{OrdinaryDiffEq.jl} package,
the runtime is 73~\% longer for $N=\num{1600}$,
6~\% shorter for $N=\num{6400}$ and 9~\% shorter for $N=\num{25600}$,
which suggests that \emph{OrdinaryDiffEq.jl} solvers are more performant
than \emph{SciPy} solvers, although for smaller resolution $N=\num{1600}$
there is clear overhead of using \emph{Open Interfaces}.
Longer runtimes for the Python code,
relative to the C and Julia codes used in Comparisons~2 and~3,
were influenced by the performance of the Python interpreter.

Overall, we can see from these three comparisons that the performance
penalty of \emph{Open Interfaces} is on average less than 5~\%
for significant workloads.

Lack of a Gateway and Convert components in Julia
at the current stage of the project has prohibited us
from conducting a comparison for Julia user codes,
however, we expect that the results of such a study would not change
the results demonstrated in Table~\ref{tab:results:burgers:runtimes}.

\begin{table}[htbp]
  \caption{%
    Run times, in seconds, of time integration of system~\eqref{eq:results:burgers:mol}
    using different user languages:
    C, Julia, or Python, with ``Julia (C)'' meaning that RHS implementation is in C),
    different ways of invoking implementations:
    via \emph{Open Interfaces} (OIF) or directly (RAW),
    and three different implementations
    (\texttt{DOPRI5-C}---C translation of the original Fortran code~\citet{HairerEtAl1993},
    \texttt{DP5} from Julia's \emph{OrdinaryDiffEq.jl} package,
    \texttt{DOPRI5}---Python wrapper over the original Fortran code~\citet{HairerEtAl1993}
    from \emph{SciPy}).%
    \label{tab:results:burgers:runtimes}
  }
  \newcommand{\myheader}{%
    \headerWithBreaks{0.2em}{\#} &
    \headerWithBreaks{3em}{User\\language} &
    \headerWithBreaks{1.5em}{OIF/\\RAW} &
    \headerWithBreaks{3em}{Imple-\\mentation} &
                                    \multicolumn{3}{c}{$N$} \\
                                    \cmidrule(lr){5-7}
  }
  \centering
  \begin{adjustbox}{max width=\textwidth}
    \begin{tabular}{l l l l c c c}
      \toprule
      \myheader                                                                                        \\
                        &           &     &          & 1600          & 6400          & \num{25600}    \\
      \midrule
      \multirow{2}{*}{1} & C         & OIF & DOPRI5-C & 0.068 ± 0.001 & 1.011 ± 0.017 & 21.006 ± 0.100 \\
                        & C         & RAW & DOPRI5-C & 0.069 ± 0.001 & 0.951 ± 0.012 & 20.699 ± 0.121 \\
      \addlinespace
      \multirow{3}{*}{2} & C         & OIF & DP5      & 0.082 ± 0.000 & 0.847 ± 0.003 & 20.700 ± 0.049 \\
                        & Julia (C) & RAW & DP5      & 0.056 ± 0.002 & 0.820 ± 0.008 & 20.364 ± 0.073 \\
                        & Julia     & RAW & DP5      & 0.067 ± 0.009 & 0.868 ± 0.004 & 21.058 ± 0.067 \\
      \addlinespace
      \multirow{3}{*}{3} & Python    & RAW & DOPRI5   & 0.113 ± 0.000 & 1.573 ± 0.010 & 30.829 ± 0.121 \\
                        & Python    & OIF & DOPRI5   & 0.122 ± 0.009 & 1.575 ± 0.005 & 30.944 ± 0.122 \\
                        & Python    & OIF & DP5      & 0.196 ± 0.003 & 1.466 ± 0.005 & 28.147 ± 0.040 \\
      \bottomrule
    \end{tabular}
  \end{adjustbox}
\end{table}

\section{Software metadata}

\subsection{Operating system}
This package does not impose any requirements on an operating system
but requires a Unix-like environment.

\subsection{Programming language}
Package building requires a~C17+ compiler.
During runtime, Python 3.11-3.13 and Julia 1.11+ are needed
for corresponding implementations.

\subsection{Dependencies}
All mandatory dependencies include a C compiler, CMake, Python, NumPy, and Julia.
We provide an environment file for the Conda package manager
for Python- and C-level dependencies,
however, users are free to use their package manager of choice instead.
Besides, Julia dependencies are installed via the official \texttt{juliaup}
utility and the built-in package manager.

\subsection{Software location:}

\textbf{Archive}

\begin{description}
	\item[Name:] MaRDI Open Interfaces
	\item[Persistent identifier:] \url{https://doi.org/10.5281/zenodo.13753666}
	\item[Licence:] BSD-2-Clause
	\item[Publisher:] Dmitry I.\ Kabanov
	\item[Version published:] 0.6.2
	\item[Date published:] 30 September 2025
\end{description}

\textbf{Code repository}

\begin{description}
	\item[Name:] MaRDI Open Interfaces
	\item[Identifier:] \url{https://github.com/MaRDI4NFDI/open-interfaces}
	\item[Licence:] BSD-2-Clause
	\item[Date published:] 30 September 2025
\end{description}

\subsection{Language}

English

\section{Reuse potential}
The software package has high reuse potential as it aims to work on general
numerical problems.
Its usage would be optimal for projects which demand integration
of differential equations or benchmarking software packages.
Although in the current state the project has limited functionality,
we plan to expand its features in the future
in terms of supported programming languages and interfaces,
such as interfaces to optimization solvers
and solvers for partial differential equations.

Contributors can participate in extending/improving this software package
in three different aspects.
First,
by adding new languages (C++, R, Rust, etc.) on the user or implementation side.
Second, by adding new implementations by writing them from scratch
according to the \texttt{IVP} interface
or by adapting existing implementations to the \texttt{IVP} interface.
Third, by adding new interfaces and implementations for other numerical problems
such as optimization.

To add support for a new language
on the user's side, one needs to write the converting function
from native data types to the C representation
and invocation of the C functions provided by the library
for loading, calling, and unloading implementations.
To add a language on the implementation side,
one needs to write the converting function
from the C representation to native data types
and implementing function for loading implementations (as modules,
shared libraries, etc.), calling their functions and unloading them.

To add support for a new problem type, one needs to define an interface
for the problem type on both the user and implementation sides,
and adapt existing implementations to the interface or write an implementation
from scratch.

Users of the software package may report issues and ask questions
using the Issues facilities of the software repository on GitHub.
or directly at \url{mailto:dmitry.kabanov@uni-muenster.de}.

\section*{Funding statement}

This work was funded by the National Research Data Infrastructure,
project number~460135501, NFDI~29/1
“MaRDI – Mathematical Research Data Initiative
[Mathematische Forschungsdateninitiative]”~\citep{BennerEtAl2022}
and additionally by Deutsche Forschungsgemeinschaft
(DFG, German Research Foundation)
under Germany's Excellence Strategy EXC~2044-390685587,
\emph{Mathematics Münster: Dynamics--Geometry--Structure}.

\section*{Acknowledgements}
Authors would like to aknowledge the help of the help of the users of the forum
\url{https://discourse.julialang.org}, particularly,
Prof.\ Steven G.\ Johnson, MIT, USA,
with optimizing the Julia implementation of the right-hand side function
for performance studies.

\section*{Competing interests}

The authors declare that they have no competing interests.

\section*{References}

\printbibliography[heading=none]

\end{document}